\documentclass[journal]{IEEEtran}
\usepackage{cite}
\usepackage{amsmath,amssymb,amsfonts}
\usepackage{algorithmic}
\usepackage{graphicx,color}
\usepackage{textcomp}
\usepackage{xcolor}
\usepackage{hyperref}
\hypersetup{hidelinks=true}
\usepackage{multirow}
\usepackage{algorithm,algorithmic}
\usepackage{tikz}
\usepackage{pgfplots}
\usepackage[caption=false,font=footnotesize]{subfig}

\newcommand{\expjwT}[1]{e^{-j\omega T #1}}
\newcommand{\exppjwT}[1]{e^{j\omega T #1}}

\newcommand{\intovertwopi}[1]{\int_{-\pi}^{\pi} #1 d(\omega T)}

\newcommand{\transpose}{^{\text{T}}}

\usepackage{fancyhdr}
\fancypagestyle{plain}{
  \fancyhf{}
  
  \fancyhead[C]{\scriptsize This work has been submitted to the IEEE for possible publication. 
  Copyright may be transferred without notice, after which this version may no longer be accessible.}
}
\begin{document}

\title{Closed-Form Least-Squares Design of Fast-Convolution Based Variable-Bandwidth FIR Filters}

\author{\IEEEauthorblockN{Oksana Moryakova,~\IEEEmembership{Graduate Student Member,~IEEE,}  and H\aa kan Johansson,~\IEEEmembership{Senior Member,~IEEE}}\\


\thanks{O. Moryakova and H. Johansson are with the Division of Communication Systems, Department of Electrical Engineering, Link\"{o}ping University, Link\"{o}ping, Sweden (e-mail: oksana.moryakova@liu.se; hakan.johansson@liu.se). \textit{Corresponding authors: O. Moryakova and H. Johansson.}} }


\maketitle
\thispagestyle{plain}
\begin{abstract}
This paper introduces a closed-form least-squares (LS) design approach for fast-convolution (FC) based variable-bandwidth (VBW) finite-impulse-response (FIR) filters. The proposed LS design utilizes frequency sampling and the VBW filter frequency-domain implementation using the overlap-save (OLS) method, that together offer significant savings in implementation and online bandwidth reconfiguration complexities. Since combining frequency-domain design and OLS implementation leads to a linear periodic time-varying (LPTV) behavior of the VBW filter, a set of the corresponding time-invariant impulse responses is considered in the proposed design. Through numerical examples, it is demonstrated that the proposed approach enables not only closed-form design of FC-based VBW filters with substantial complexity reductions compared to existing solutions for a given performance, but also allows the variable bandwidth range to be extended without any increase in complexity. Moreover, a way of reducing the maximum approximation error energy over the whole set of the time-invariant filters of the LPTV system is shown by introducing appropriate weighting functions in the design.
\end{abstract}


\section{Introduction}
\label{sec:introduction}
\IEEEPARstart{D}{ue} to the increasing demand for reconfigurable systems in the contemporary world of technologies, variable digital filters (VDFs) are required in many digital signal processing applications, for example, in medical devices \cite{Indrakanti_2018} and communication systems \cite{RAGHU19, Renfors_2014, Canese_2023}. The main advantage of these filters over regular digital filters is that they offer variability of the frequency response by adjusting only one or a few parameters without online filter design. Most recent papers focused on VDFs \cite{Indrakanti_2018, RAGHU19, Renfors_2014, Canese_2023, Sarband_2025} have shown that this approach allows to significantly reduce implementation complexity and thereby hardware complexity compared to regular filters requiring online design for every new specification. Nevertheless, implementations of VDFs in the time domain may still cause a rather high computational complexity for stringent requirements.

Research works on finite-impulse-response (FIR) filters have shown that a filter can be implemented in the frequency domain using the fast convolution (FC) with much lower complexity than in the time domain, especially using the overlap-save (OLS) technique \cite{Shynk_1992, Ishihara_2011, Kovalev_2017, Lin_2022, Renfors_2014, Renfors_2018, Yli-Kaakinen_2022, Moryakova_2023}.
The FC-based VDFs have shown significant reduction of the computational complexity compared to time-domain implementations \cite{Renfors_2014, Lin_2022, Renfors_2018, Yli-Kaakinen_2022, Moryakova_2023}. 

Typically, regardless of the implementation, VDFs are designed by optimizing the impulse response values to satisfy specification requirements for the frequency response, hereafter this is also referred as \textit{time-domain design}. In this case, the filter's discrete Fourier transform (DFT) coefficients are given by the DFT of the optimized impulse response. 
Since the time-domain design does not take into account the frequency-domain implementation aspects, the complexity of implementing such filters in the frequency domain can be relatively high, that leads to inefficient implementation, especially for an update of variable filter characteristics referred as \textit{online reconfiguration} \cite{Moryakova_2023}.

To overcome this problem, one option is to directly use the DFT coefficients as design parameters, hereafter this is referred as \textit{frequency-domain design}. This approach, which is typically combined with frequency sampling \cite{Rabiner_1970, Harris_1998, SalcedoSanz_2007, Belorutsky_2016, Renfors_2014, Moryakova_2024_HFSO}, together with the OLS-based implementation causes the resulting filter to behave as a linear periodically time-varying (LPTV) system \cite{Renfors_2014}.
This complicates the design process, as the performance must be evaluated by using distortion and aliasing functions, typically analyzed through a multirate filter bank model, or via a periodically time-varying impulse response (PTVIR) representation using a set of corresponding time-invariant responses \cite{Mehr_02, Vaidyanathan_MSFB, Johansson_23}.
For regular FIR filters, the frequency-domain design approach based on frequency sampling has been studied \cite{Rabiner_1970, Harris_1998, SalcedoSanz_2007, Belorutsky_2016}, but the authors did not consider filter implementations in the frequency domain, i.e., they did not analyze the LPTV behavior.
In contrast, a multirate FC filter bank design approach in \cite{Renfors_2014} utilized the fact that a frequency-domain designed FC-based filter, causing cyclic distortions, can be modeled as an LPTV system,
but the authors did not provide details whether the corresponding time-invariant responses have been controlled during the design procedure and they did not consider VBW filters.
For FC-based VBW filters, an approach based on frequency sampling and minimax optimization has been introduced in \cite{Moryakova_2024_HFSO}. This method considers the use of the same transition band DFT coefficients for different bandwidths in the OLS implementation and takes into account the LPTV behavior by controlling the corresponding set of time-invariant responses in the design. 
Although the approach in \cite{Moryakova_2024_HFSO} significantly reduces the complexity of implementation and online reconfiguration compared to existing solutions \cite{Lowenborg_2006, Moryakova_2023}, the design method relies on iterative optimization, which is computationally demanding and time-consuming, especially when there are many constraints, as for VBW filters.

In this work, utilizing the idea in \cite{Moryakova_2024_HFSO} of using the same set of transition band coefficients for different bandwidths, we introduce a closed-form design approach for FC-based VBW filters optimum in the least-squares (LS) sense. It is based on the work in \cite{Moryakova_2024_HFSO} presented at a conference by the authors.
The main contributions of the paper are as follows.
\begin{enumerate}
	\item  The proposed design method is derived in the LS sense combining frequency sampling and optimization, and explicitly accounting for the LPTV nature of the OLS implementation using a set of time-invariant impulse responses. As a result, the method eliminates the need for iterative optimization as in \cite{Moryakova_2024_HFSO}, providing a closed-form expression for obtaining the filter DFT coefficients. Additionally, since the passband and stopband values are obtained directly by sampling the desired frequency response, and only $K$ transition-band values need to be optimized, the offline design complexity of the proposed LS approach simplifies to a $K\times K$ matrix inversion, i.e., it is reduced compared to \cite{Moryakova_2024_HFSO}.
	\item The implementation and online reconfiguration complexities of the VBW filter designed using the proposed approach are reduced and do not depend on the variable bandwidth range, while it is one of the main limitations in existing approaches \cite{Lowenborg_2006, Moryakova_2023}. Since the proposed method considers that the magnitude of the filter DFT coefficients for various bandwidths are constructed based on ones for the passband, zeros for the stopband, and the optimized transition band values, this eliminates any computations when the bandwidth is varied and leads to a very simple update of the filter DFT coefficients.
	\item A systematic design procedure for FC-based VBW filters with arbitrary specifications is provided. This is an extension of \cite{Moryakova_2024_HFSO}, which considers only specifications that match DFT frequency bins.
	\item Through numerical examples, we demonstrate that the proposed method achieves performance comparable to existing optimization-based approaches \cite{Lowenborg_2006, Moryakova_2023, Moryakova_2024_HFSO}, while offering complexity reductions. These features make the method attractive for applications requiring efficient VBW filters with closed-form design. 
	\item Furthermore, a refinement of the LS formulation through appropriate weighting functions is presented. By utilizing the LPTV properties, it reduces the maximum approximation error energy across the set of equivalent time-invariant filters.
\end{enumerate}

The rest of the paper is organized as follows. Following Introduction, Section \ref{sec:OLS} gives a brief overview of the OLS method. Further, Section \ref{sec:prop_design} presents the proposed closed-form design of an FC-based VBW filter, while Section \ref{sec:examples} provides numerical examples and computational-complexity analysis. Finally, Section \ref{sec:conclusion} concludes the paper.

\section{Overlap-Save Method}
\label{sec:OLS}
The main concept of the FC using the OLS method is that the input signal  is divided into overlapping segments $x_m(n)=x(n+mM)$, $n=0,1,..., N-1$, where $m$ is a segment index and the overlapping part is $N-M$. Then, for each segment, the following computations are carried out.
\begin{enumerate}
	\item The segment $x_m$ is transformed via an $N$-point DFT.
	\item The DFT coefficients $X_m(k)$ are multiplied by the filter DFT coefficients $H(k)$, $k=0,1,\dots,N-1$.
	\item An $N$-point inverse DFT (IDFT) is performed.
	\item The first $N-M$ samples of the resulting block are discarded, so that the output segments of length $M$ are no longer overlapping. 
	\item The output sequence is obtained by concatenating the resulting segments as $y(n)=\sum_{m=0}^{\infty}y_m(n-mM)$.
\end{enumerate}

In classical OLS filtering, the system is intended to be time-invariant, i.e., the aliasing error is zero. In this case, the  DFT coefficients $H(k)$ of an FIR filter correspond to the $N$-point DFT of the impulse response $h(n)$ of length $L$, designed by any of the available methods \cite{Oppenheim_DTSP}. This means $h(n)=0$ for $n=L,...,N-1$, where $L=N-M+1$.

In this paper, the proposed design approach, which will be discussed in details in Section \ref{sec:prop_design}, is based on combining frequency sampling and filter DFT coefficients optimization, that together restricts the frequency response to take on fixed values at $N$ uniformly spaced frequencies. Applying an $N$-point IDFT to these samples yields an impulse response of length~$N$. In contrast to the classical OLS method, with only the first $L$ samples of the impulse response being non-zero, this approach results in a full-length impulse response with all $N$ values generally non-zero. Although the effective order of the underlying filter is approximately $L-1$ and the impulse response values for $n\in[L, N-1]$ are typically small in magnitude, their effect is non-negligible. Specifically, due to the block-wise nature of the OLS implementation, the overall system becomes $M$-periodic time-varying, i.e., the filter impulse response coefficients vary from sample to sample with period~$M$ \cite{Daher_2010, Renfors_2014, Johansson_23}.
In these systems, aliasing cannot be cancelled but can be suppressed to any desired level through a proper design, which is considered in this paper. 

\section{Proposed Closed-Form Design of a VBW Filter Implemented Using the OLS Method}
\label{sec:prop_design}
\vspace{3pt}
\subsection{VBW Filter Implemented via the OLS method}
\label{subsec:VBW_filt}
The desired frequency response of a VBW filter of effective length  $L$ is considered here as
\begin{equation}
	H_{D_1}(\exppjwT{}, b) = 
	\begin{cases}
		\expjwT{D_1},  & \omega T \in[0, b-\Delta/2],\\
		0,  & \omega T \in[b+\Delta/2, \pi],
	\end{cases}
	\label{eq:VBW_des}
\end{equation}
where the variable parameter $b\in[b_l, b_u]$ is the the center of the transition band, whereas  $b-\Delta/2$ and $b+\Delta/2$ represent passband and stopband edges, respectively. The transition width $\Delta$ and delay $D_1=(L-1)/2$ are assumed to be fixed.

\begin{figure}
	\centering
	\includegraphics[width=\linewidth]{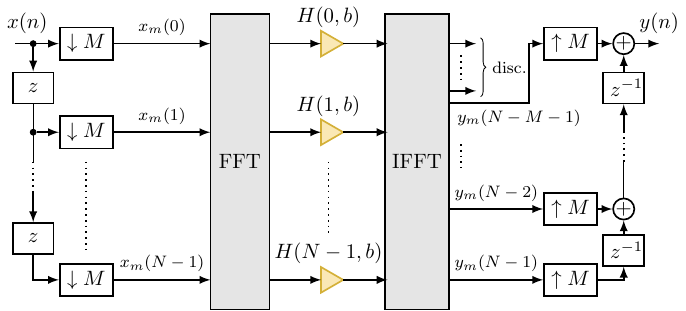}
	\caption{Frequency-domain implementation of a VBW filter using the overlap-save method \cite{Moryakova_2023}.}
	\label{fig:VBW_FD_original}
\end{figure}

The VBW filter frequency-domain implementation with the lowest implementation complexity proposed in \cite{Moryakova_2023} is shown in Fig. \ref{fig:VBW_FD_original}. This structure utilizes $N$ variable DFT coefficients $H(k,b)$, $k=0,1,\dots,N-1$, that have to be recomputed whenever $b$ is changed. To reduce this online reconfiguration complexity, specifically, to eliminate any computations for a new $b$-value, the VBW filter in Fig. \ref{fig:VBW_FD_original} is designed using the frequency-domain approach in this paper.  

As was mentioned in Section \ref{sec:OLS}, a filter designed using frequency sampling and implemented using the OLS method, becomes an LPTV system.
In general, the behavior of an LPTV system can be analyzed by representing it in terms of its PTVIR, which corresponds to a set of time-invariant impulse responses $h_n(q)=h_{n+M}(q)$, 
$n = 0, 1,..., M-1$, $q = 0, 1, ..., N-1$, and their corresponding frequency responses $H_n(\exppjwT{})$ \cite{Daher_2010, Vaidyanathan_MSFB, Johansson_23}. This representation can be a better indicator of the worst-case time-domain error of the overall system comparing to the multirate filter-bank representation in terms of distortion and aliasing \cite{Johansson_23}. 
Thus, the implementation in Fig. \ref{fig:VBW_FD_original} can be equivalently analyzed using the PTVIR representation, and the output of the system can be written as
\begin{equation}
	y(n, b)=\frac{1}{2\pi}\intovertwopi{H_n(\exppjwT{}, b)X(\exppjwT{})\exppjwT{n}},
\end{equation}
where $H_n(\exppjwT{}, b)$ are the frequency responses, as derived in \cite{Johansson_23} for a regular FIR filter, but expressed here with the inclusion of the variable parameter $b$. Following an analysis of the impulse response properties for the OLS method in \cite{Johansson_23} and extending it for VBW filters, the responses $H_n(\exppjwT{}, b)$ can be expressed as
\begin{equation}
	H_n(\exppjwT{}, b)=\expjwT{n}\sum_{q=0}^{N-1}d_n(q, b)\expjwT{q}
	\label{eq:OLS_Hn}
\end{equation}
with $d_n(q,b)$ being the impulse responses of $z^n H_n(z, b)$ and circular versions of each other, i.e.,
\begin{equation}
	d_{n+m}(q, b)=d_n((q+m)\text{ mod } N, b).
	\label{eq:dn_circ}
\end{equation}
This allows to avoid computations of $d_n(q,b)$ for every $n$ in the design. 
Then the impulse responses $h_n(q,b)$ can be seen as responses $d_n(q,b)$ delayed by $n$ samples, specifically
the last response $h_{M-1}(q,b)$ corresponds to the delayed $d_{M-1}(q,b)$ by $M-1$ samples, where $d_{M-1}(q,b)$ is the IDFT of $H(k,b)$ \cite{Johansson_23}, i.e.,
\begin{equation}
	d_{M-1}(q, b) = \frac{1}{N}\sum_{k=0}^{N-1}H(k, b)e^{j2\pi qk/N}.
	\label{eq:OLS_dn}
\end{equation}
Thus, using \eqref{eq:dn_circ} and \eqref{eq:OLS_dn}, all the responses $d_n(q, b)$ can be expressed in terms of $d_{M-1}(q, b)$ as
\begin{align}
	d_n(q, b) &= 
	d_{M-1}((q+n-M+1) \text{ mod } N, b),
	\label{eq:dn_dM-1}
\end{align}
and $h_n(q,b)$ can further be obtained based on $d_{M-1}(q)$ as
\begin{align}
	\hspace{-5pt}h_n(q,b) = 
	\begin{cases}
		0,  & q<n,\\
		d_{M-1}((q+P) \text{ mod } N,b), &n\leq q < N,\\
		0,  & \text{otherwise},
	\end{cases}
\end{align}
where $P=n-M+1$, $n = 0, 1,..., M-1$. As an illustration of the circular property for this LPTV system, the impulse response samples $h_n(q,b)$ are listed in terms of $d_{M-1}(q,b)$ in Table \ref{tab:dn_responses} for $N=8$ and $M=4$. 

Further, the DFT coefficients $H(k,b)$ in \eqref{eq:OLS_dn} are obtained using frequency sampling and filter DFT coefficient optimization as will be described in the following subsection.

\begin{table}[t]
	\caption{Impulse Responses $h_n(q,b)$ for $N=8$ and $M=4$ \\Showing the Circular Property of the Responses $d_n(q, b)$}
	\begin{center}
		\begin{tabular}{c c c c}
			\hline
			$h_0(q, b)$ & $h_1(q, b)$ & $h_2(q, b)$ & $h_3(q, b)$\\
			\hline
			$d_3(5,b)$  &  0          	 &  0          		&  0\\
			$d_3(6,b)$  & $d_3(6,b)$   &  0          	  &  0\\
			$d_3(7,b)$  & $d_3(7,b)$  &  $d_3(7,b)$ &  0\\
			$d_3(0,b)$  &  $d_3(0,b)$ &  $d_3(0,b)$ &  $d_3(0,b)$\\
			$d_3(1,b)$  &  $d_3(1,b)$ &  $d_3(1,b)$ &  $d_3(1,b)$\\
			$d_3(2,b)$  &  $d_3(2,b)$ &  $d_3(2,b)$ &  $d_3(2,b)$\\
			$d_3(3,b)$  &  $d_3(3,b)$ &  $d_3(3,b)$ &  $d_3(3,b)$\\
			$d_3(4,b)$  &  $d_3(4,b)$  &  $d_3(4,b)$ &  $d_3(4,b)$ \\
			0                &  $d_3(5,b)$ &  $d_3(5,b)$ &  $d_3(5,b)$\\
			0           	 &  0          	 	& $d_3(6,b)$ 	 & $d_3(6,b)$\\
			0           	 &  0          		&  0          	   &  $d_3(7,b)$\\
			\hline
		\end{tabular}
	\end{center}
	\label{tab:dn_responses}
\end{table}

\subsection{Proposed Approach to Closed-Form Design}
\label{subsec:LS_design}
As mentioned in Section \ref{subsec:VBW_filt}, frequency-domain implementations of a VBW filter designed in the frequency domain can be analyzed using a set of $M$ periodic time-invariant responses $H_n(\exppjwT{}, b)$. Therefore, assuming real-valued filter impulse responses and even\footnote{For complex-valued responses  and/or odd DFT lengths, a similar design strategy can be used after some minor modifications.} $N$, the proposed approach to design a VBW filter consists of three steps:
\begin{enumerate}
	\item[(i)] Sampling the desired frequency response $H_{D_1}(\exppjwT{},b)$ in \eqref{eq:VBW_des} and obtaining the DFT coefficients $H(k,b)$ in the passband and stopband regions. For the transition band, where the values are generally not specified, the sample indices of the DFT coefficients are defined in terms of the variable parameter $b$.
	\item[(ii)] Expressing the $M$ time-invariant responses $H_n(\exppjwT{}, b)$ based on the known passband DFT coefficients and unknown transition band DFT coefficients from the previous step using \eqref{eq:OLS_Hn}–\eqref{eq:OLS_dn}.
	\item[(iii)] Minimizing the error energy among all $M$ responses and the desired response for $b\in[b_l, b_u]$ by optimizing the transition band DFT coefficients.
\end{enumerate}

\subsubsection*{Step I – Sampling the Desired Response}
\begin{figure}[!t]
	\centering
	\subfloat[Sampling the desired response of the VBW filter in \eqref{eq:VBW_des}.]{
		\includegraphics[trim={4.3cm 5.5cm 7.8cm 9.3cm},clip,width=0.95\linewidth]{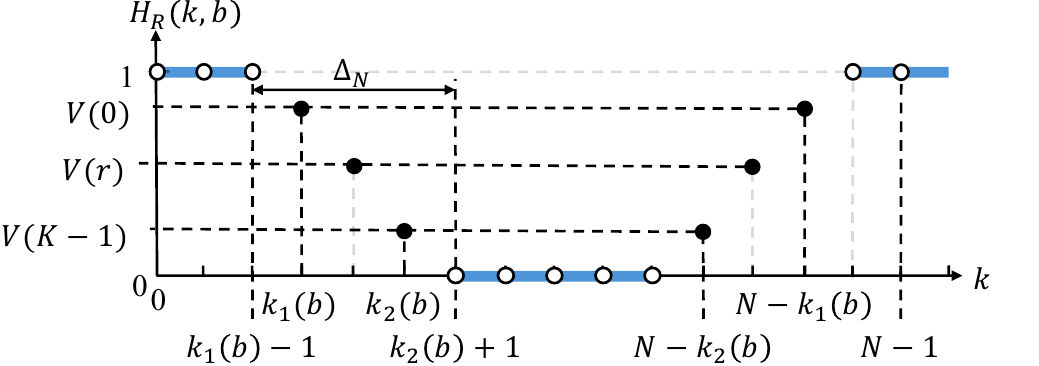}%
		\label{fig:Sampling_VBW}
	} \\[1ex]
	\subfloat[Reusing the same set of transition band samples for different $b$-values.]{
		\includegraphics[trim={4.3cm 5.4cm 7.8cm 7.7cm},clip,width=0.95\linewidth]{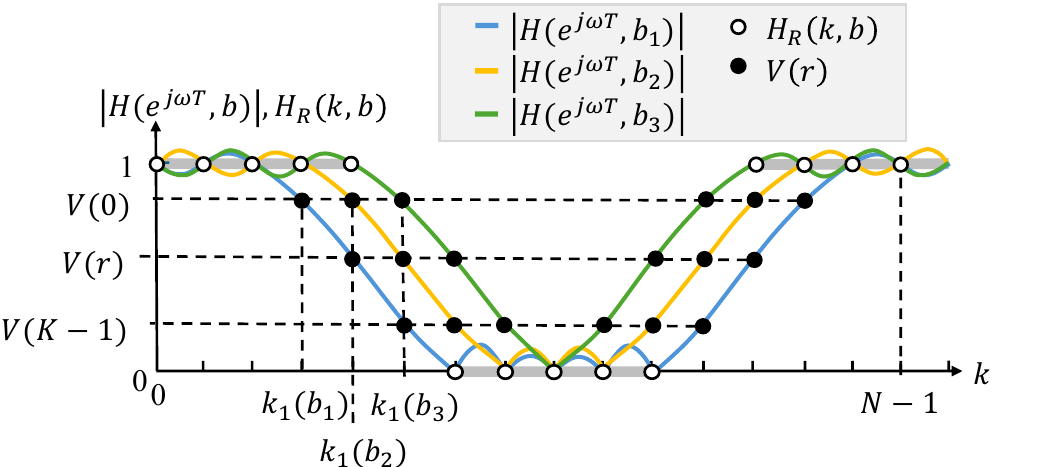}%
		\label{fig:Sampling_VBW_ex}
	}
	\caption{Examples of sampling and reuse in the VBW filter design.}
	\label{fig:VBW_examples}
\end{figure}

By sampling the desired response in \eqref{eq:VBW_des}, we set the DFT coefficients $H(k,b)$ in the form  
\begin{equation}
	H(k,b) = H_R(k,b)e^{-j\frac{2\pi k D_1}{N}},
	\label{eq:FS_Hk_exp}
\end{equation}
with the magnitude response samples $H_R(k,b)$ given by 
\begin{align}
	H_R(k,b) =
	\begin{cases}
		1,  & k\in \textbf{k}_{p}(b),\\
		V(k-k_1(b)),  & k\in \textbf{k}_t(b),\\
		0,  & k\in \textbf{k}_s(b),
	\end{cases}
	\label{eq:FS_Hk_real}
\end{align}
where the passband, transition band, and stopband regions are defined, respectively, by
\begin{subequations}
	\begin{align}
		\textbf{k}_p(b)&= [0,k_1(b)-1] \cup [N-k_1(b)+1,N-1],\\
		\textbf{k}_t(b)&=[k_1(b), k_2(b)] \cup [N-k_2(b), N-k_1(b)],\\
		\textbf{k}_s(b)&=[k_2(b)+1, N-k_2(b)-1].
	\end{align}
\end{subequations}
The form in \eqref{eq:FS_Hk_exp} allows the filter to have an approximately linear phase after sampling.
Illustration of the frequency sampling of the VBW frequency response is shown in Fig.~\ref{fig:Sampling_VBW}. Here, $k_1(b)$ and $k_2(b)$ define the first and the last samples of the transition band, correspondingly, given by
\begin{subequations}
	\begin{align}
		k_1(b)&=b_N(b)-\Delta_N/2+1,\label{eq:k_var1}\\
		k_2(b)&=b_N(b)+\Delta_N/2-1=k_1(b)+K-1,
		\label{eq:k_var}
	\end{align}
\end{subequations}
where $K$ is the number of the transition band samples, $\Delta_N$ is the fixed transition width in terms of frequency bins as
\begin{align}
	\Delta_N=k_2(b)-k_1(b)+2=K+1,
\end{align}
and $b_N(b)\in[b_{N}(b_l), b_{N}(b_u)]$ is the variable parameter in terms of frequency bins, $\Delta_N/2\leq b_{N}(b_l) < b_{N}(b_u) \leq N/2-\Delta_N/2-1$. 
The values $k_1(b)$, $k_2(b)$, $\Delta_N$, and $b_N(b)$ are obtained based on the desired filter specification and will be discussed in Section \ref{subsec:procedure_complexity}.

In \eqref{eq:FS_Hk_real}, it is assumed that the set of $K$ transition band samples $V(k-k_1(b))$, $k\in\textbf{k}_t$, can be utilized for different values of $b$ as it is illustrated in Fig. \ref{fig:Sampling_VBW_ex}. Therefore, the values of the set do not depend on $b$ and can be written as $V(r)$, $r=0,\dots, K-1$.

\subsubsection*{Step II – Expressing Time-Invariant Responses}
Considering that the VBW filter impulse response is real-valued, the responses $d_n(q,b)$ are also real-valued. Thus, using \eqref{eq:OLS_dn} and \eqref{eq:FS_Hk_exp}, the coefficients $H(k,b)$ are conjugate-symmetric, and $d_{M-1}(q,b)$ can be expressed as
\begin{multline}
	d_{M-1}(q,b)=\frac{1}{N}\sum_{k=0}^{N-1}H_R(k,b)\cos\Bigg(\frac{2\pi k}{N}\big(D_1-q\big)\Bigg)\\
	=\frac{1}{N}\Bigg[1+2\sum_{k=1}^{N/2-1}H_R(k,b)\cos\Bigg(\frac{2\pi k}{N}\big(D_1-q\big)\Bigg)\Bigg].
\end{multline}
Further, introducing a function $f(k,q,n)$ given by
\begin{align}
	f(k,q,n) = \cos\Big(\frac{2\pi k}{N}\big(D_2-(q+n)\big)\Big),
\end{align}
with
\begin{align}
	D_2=D_1+M-1,
	\label{eq:delay_D2}
\end{align}
and using \eqref{eq:dn_dM-1} and \eqref{eq:FS_Hk_real},
$d_n(q,b)$ can be represented as
\begin{multline}
	d_{n}(q,b) = \frac{1}{N}\bigg[1+2\sum_{k=1}^{k_1(b)-1}f(k,q,n)\\ +2\sum_{k=k_1(b)}^{k_2(b)}V(k-k_1(b))f(k,q,n)\bigg].
	\label{eq:dn_IDFT_real}
\end{multline}

Then, inserting \eqref{eq:dn_IDFT_real} in \eqref{eq:OLS_Hn} and replacing $V(k-k_1(b))$ with $V(r)$ as mentioned in Step I, $H_n(\exppjwT{}, b)$ can be written as
\begin{align}
	\hspace{-4pt}
	H_n(\exppjwT{}, b)  
	=\frac{1}{N}&\sum_{q=0}^{N-1}\expjwT{(q+n)}
	\bigg[1
	+ 2\sum_{k=1}^{k_1(b)-1}f(k,q,n)\notag\\
	&+ 2\sum_{r=0}^{K-1}V(r)f(r+k_1(b),q,n)\bigg].
	\label{eq:Hn_real}
\end{align}

\subsubsection*{Step III – Optimizing Transition Band Values}
The problem of designing a VBW filter implemented using the OLS method is to find the transition band coefficients $V(r)$, $r=0,1, \dots, K-1$, such that for every $n=0, 1, \dots, M-1$ the response $H_n(e^{j\omega T}, b)$ approximates the desired response $H_{D_2}(e^{j\omega T}, b)$  given by
\begin{equation}
	H_{D_2}(\exppjwT{}, b) = 
	\begin{cases}
		\expjwT{D_2}, \quad & \omega T\in [0, b-\Delta/2], \\
		0, \quad & \omega T\in [b+\Delta/2, \pi]
	\end{cases}
	\label{eq:OLS_Hd}
\end{equation}
for $b\in[b_l, b_u]$ and with $D_2$ as in \eqref{eq:delay_D2}, which is different from $D_1$ in \eqref{eq:VBW_des} by additional $M-1$ samples due to the OLS implementation as it can be seen in Fig. \ref{fig:VBW_FD_original}.
To obtain the coefficients $V(r)$ analytically, the approximation problem is formulated in the LS sense. Specifically, one needs to minimize the error function $E$, which includes all $M$ responses and the variable range of $b$-values in terms of frequency bins as mentioned in Step I, given by
\begin{multline}
	E = \sum_{b=b_{N}(b_l)}^{b_{N}(b_u)} \sum_{n=0}^{M-1} \frac{1}{2\pi}\int_{\Omega(b)} \big|H_n(\exppjwT{},b)\\
		-H_{D_2}(\exppjwT{},b)\big|^2 d(\omega T),
	\label{eq:loss_fun}
\end{multline}
where $\Omega(b)=[0, b-\Delta/2]\cup[b+\Delta/2, \pi]$.
By substituting \eqref{eq:Hn_real} and \eqref{eq:OLS_Hd} into \eqref{eq:loss_fun}, combining terms containing $V(r)$, and performing some algebraic manipulations, the energy function can be expressed in matrix form
\begin{align}
	E=\textbf{v}\transpose \textbf{Q} \textbf{v} + 2\textbf{v}\transpose \textbf{c} + c_0,
	\label{eq:loss_fun_matrix}
\end{align}
where $\textbf{v}=[V(0), V(1), ..., V(K-1)]\transpose$ is a vector containing the transition band coefficients, $c_0$ is a term independent of the coefficients $V(r)$, $\textbf{Q}$ is a $K\times K$ symmetric and positive-definite matrix\footnote{This can be shown in a similar way as for a regular FIR filter design \cite{Vaidyanathan_MSFB}.} with entries $Q(r,s)$ given by
\begin{multline}
	Q(r,s)=\frac{4}{N^2}\sum_{b=b_{N}(b_l)}^{b_{N}(b_u)}\sum_{n=0}^{N-1} \sum_{q=0}^{N-1}f(r+k_1(b),q,n)\\
	\times\sum_{p=0}^{N-1}I_1(q,p)
	f(s+k_1(b),p,n),
	\label{eq:Q_elem}
\end{multline}
where 
\begin{align}
	\hspace{-0.2cm}I_1(q,p)=
	\begin{cases}
		\frac{1}{\pi}(\pi-\Delta),  &\hspace{-0.3cm}p=q,\\
		\frac{1}{\pi}\frac{-2\sin\big(\Delta(q-p)/2\big)\cos\big(b(q-p)\big)}{q-p}, &\hspace{-0.3cm}\text{otherwise}.
	\end{cases}
\end{align}
Further, $\textbf{c}$ represents a length-$K$ vector  with elements $c(s)$ given by
\begin{multline}
	c(s)
	=\frac{2}{N}\sum_{b=b_{N}(b_l)}^{b_{N}(b_u)}\sum_{n=0}^{N-1}
	\sum_{p=0}^{N-1}f(s+k_1(b),p,n)\\ \hspace{-0.2cm}\times\bigg[\frac{1}{N}\sum_{q=0}^{N-1}I_1(p,q)
	\Big(1+2\hspace{-0.2cm}\sum_{k=1}^{k_1(b)-1}f(k,q,n)\Big)-I_2(p,n)\bigg]\hspace{-1pt},\hspace{-6.5pt}
	\label{eq:c_elem}
\end{multline}
where 
\begin{align}
	\hspace{-0.2cm}I_2(p,n)=
	\begin{cases}
		\frac{1}{\pi}(b-\Delta/2),  &\hspace{-0.2cm}p+n=D_2,\\
		\frac{1}{\pi}\frac{\sin\big((b-\Delta/2)(D_2-(p+n))\big)}{D_2-(p+n)}, &\hspace{-0.2cm}\text{otherwise}.
	\end{cases}
\end{align}
By setting the partial derivative of $E$ in \eqref{eq:loss_fun_matrix} with respect to $\textbf{v}$ to zero and solving for $\textbf{v}$, the solution to the approximation problem is obtained as
\begin{align}
	\hat{\textbf{v}}=-\textbf{Q}^{-1}\textbf{c}.
	\label{eq:LS_solution}
\end{align}

\subsection{Specification in Terms of Frequency Bins}
\label{subsec:procedure_complexity}
Since the proposed approach to design a VBW filter directly involves the values of the filter DFT coefficients as discussed in Section~\ref{subsec:LS_design}, the specification of the filter needs to be represented in terms of frequency bins for a given DFT length $N$, and the overall complexity depends on the values of $N$, $L$, and $K$, where the $K$-value is computed based on the DFT length $N$.
Thus, to design a VBW filter with the minimum overall complexity, the values of $N$ and $L$ have to be determined prior to the LS design. 

The $L$-value can be obtained as $L=N_D+1$ based on the filter order $N_D$, which can be estimated using one of the existing expressions \cite{Wang_2018, Vaidyanathan_MSFB, Bellanger_84} or taken from the minimax based designed VBW filter for comparison as will be discussed Section~\ref{sec:examples}. The optimal value of $N$ is estimated using $\widehat{N}=0.9L\log_2(L)$ \cite{Johansson_23} and rounded to the nearest $2^{Q}$. After that, the value of $M$ is computed as $M = N-L+1$.

Further, the fixed transition band width $\Delta$ can be expressed as $\Delta_N$ DFT bins given by
\begin{equation}
	\Delta_N = \lfloor\Delta\times N/(2\pi) \rfloor,
	\label{eq:Delta_N}
\end{equation} 
that corresponds to the truncated transition band width $\Delta_D$ given by
\begin{equation}
	\Delta_D = \Delta_N\times 2\pi/N,
	\label{eq:Delta_D}
\end{equation}
where $\Delta_D\leq\Delta$. This means that the specification discretized based on the $N$-value can be tightened. It is worth noting that $\Delta_N$ has to be even to obtain integer values for $k_1(b)$ and $k_2(b)$ in \eqref{eq:k_var1} and \eqref{eq:k_var}, respectively. Thus, if $\Delta_N$ in \eqref{eq:Delta_N} is odd, $\Delta_N\leftarrow\Delta_N-1$, and $\Delta_D$ in  \eqref{eq:Delta_D} is recomputed. Further, the filter order $N_D$, and correspondingly the value of $L$, must be recomputed based on the discretized $\Delta_D$. 

Similarly, in order to define the transition band edges $k_1(b)$ and $k_2(b)$ in \eqref{eq:k_var1}–\eqref{eq:k_var}, the variable parameter $b\in[b_l, b_u]$ needs to be expressed in terms of frequency bins as $b_N(b)\in[b_N(b_l), b_N(b_u)]$, where the boundary values of the variable parameter are computed as
\begin{align}
	b_N(b_l) = \lfloor b_l N/(2\pi) \rfloor,\quad
	b_N(b_u) = \lceil b_u N/(2\pi) \rceil.
\end{align}
Additionally, the value of the discretized $b$-value in terms of frequency bins needs to be rounded as 
\begin{align}
	b_N(b)=\left[b\times N/(2\pi)\right].
	\label{eq:b_N}
\end{align}
Then, based on obtained values of $\Delta_N$ and $b_N$, the indices $k_1(b)$ and $k_2(b)$ in \eqref{eq:k_var1}–\eqref{eq:k_var} defining the boundaries of the transition band can be computed. 
Further, based on the discretized specification, the matrix and vector elements in \eqref{eq:Q_elem} and \eqref{eq:c_elem}, respectively, can be constructed and the solution in \eqref{eq:LS_solution} can be obtained.

\subsection{Complexity Analysis}
\label{sec:complexity}
The overall complexity of the proposed VBW filter consists of three parts: the cost of obtaining the transition-band coefficients (offline design complexity), the implementation cost, and the cost of adjusting the filter bandwidth (online reconfiguration complexity).

\subsubsection*{Offline Design Complexity}
The offline design includes the cost of solving \eqref{eq:LS_solution}. Since the main advantage of VBW filters is that the coefficients need to be computed only once, the offline design complexity is usually not a critical factor compared to other complexities. However, comparing to the approach in \cite{Moryakova_2024_HFSO}, which relies on computationally demanding and time-consuming minimax optimization,
the proposed algorithm provides a direct solution. Furthermore, because only $K$ transition-band coefficients need to be computed, the offline design reduces to inverting a $K\times K$ matrix.

\subsubsection*{Implementation Complexity}
The implementation complexity of the frequency-domain implementations consists of DFT/IDFT transforms and multiplications by $H(k,b)$. In this paper, we consider that the former is implemented using the split-radix fast Fourier transform (FFT) algorithm and the segments $x_m$ and $y_m$ are real-valued and $N=2^Q$. Therefore, each FFT and its inverse (IFFT) require $C_{mf,F}$ multiplications and $C_{a,F}$ additions, given respectively by \cite{Sorensen_1987}
\begin{gather}
	C_{mf,F}=(1/2)N\log_2(N)-(3/2)N+2,\\
	C_{a,F}=(3/2)\log_2(N)-(5/2)N+4.
\end{gather} 

The multiplications by $H(k,b)$ can be implemented very efficiently when the order $L-1$ of the VBW filter is even. Considering an input segment $x_m$, its corresponding DFT bins $X_m(k)$, and the filter DFT coefficients $H(k,b)$ as in \eqref{eq:FS_Hk_exp}, the output of the IDFT is computed as
\begin{align}
	y(n,b)
	&=\frac{1}{N}\sum_{k=0}^{N-1}X_m(k)H_R(k,b)e^{j\frac{2\pi k}{N}\big(n-D_1\big)}.
\end{align}
For the OLS implementation, it means that the complex multiplication by the filter DFT coefficient $H(k,b)$ can be realized as a real multiplication by $H_R(k,b)$ and rotations of the output. Compared to the conventional case, where the first $L-1$ output samples are discarded, here the first $(L-1)/2$ and the last $(L-1)/2$ samples are discarded due to rotation of the output segment by $D_1$ samples. This corresponds to a symmetric implementation of the OLS method (zero-phase model \cite{Renfors_2014}). Thus, only real multiplications by $H_R(k,b)$ need to be computed.
Additionally, considering that the DFT coefficients $X_m(k)$ and $H(k,b)$ are conjugate symmetric and for unit- and zero-valued $H_R(k,b)$, i.e., outside the transition band, there are no multiplications required, the number of general multiplications needed for the multiplication by the filter coefficients are $C_{mv,H}=2K$ per $M$  output samples. 

Therefore, the total implementation complexity per output sample can be expressed as a fixed multiplication rate $R_{mf}$, general multiplication rate $R_{mv}$ (multiplications with variable coefficients), and addition rate $R_a$, given respectively by
\begin{align}
	R_{mf} &= \frac{2C_{mf, F}}{M}=\frac{N\log_2(N)-3N+4}{N-L+1},\label{eq:R_mf}\\
	R_{mv} &= \frac{C_{mv,H}}{M}=\frac{2K}{N-L+1},\label{eq:R_mv}\\
	R_a &=\frac{2C_{a,F}}{M}=\frac{3N\log_2(N)-5N+8}{N-L+1}\label{eq:R_a}
\end{align}
with $M = N-L+1$ for the OLS method. 

\subsubsection*{Online Reconfiguration Complexity} 
Since the proposed approach allows the use of the same transition band DFT coefficients for different bandwidths, only the set of $K$ values $V(r)$ needs to be stored in memory. Additionally, the discretized $b$-value is computed as in \eqref{eq:b_N} when $b$ is changed. Thus, the online reconfiguration complexity per output sample consists of $1/M$ multiplications in the worst case (if $b$ is varied for every block of $M$ samples), which is approximately zero for large $M$, and memory to store $K$ values of $V(r)$.

\section{Design Examples}
\label{sec:examples}
In this section, we apply the proposed VBW filter design and compare the resulting OLS-based implementation with 
\begin{enumerate}
	\item[(i)] a VBW filter designed in the time domain (TD) and implemented using the Farrow structure \cite{Lowenborg_2006},
	\item[(ii)] a VBW filter designed in the TD and implemented in the frequency domain (FD) using the OLS method \cite{Moryakova_2023},
	\item[(iii)] a VBW filter designed using the hybrid of frequency sampling and minimax optimization and implemented using the OLS method \cite{Moryakova_2024_HFSO}.
\end{enumerate}

\subsection{Example 1}
\begin{figure}[t]
	\includegraphics[trim={2.4cm 2.6cm 2.4cm 2.6cm}, clip,width=\linewidth]{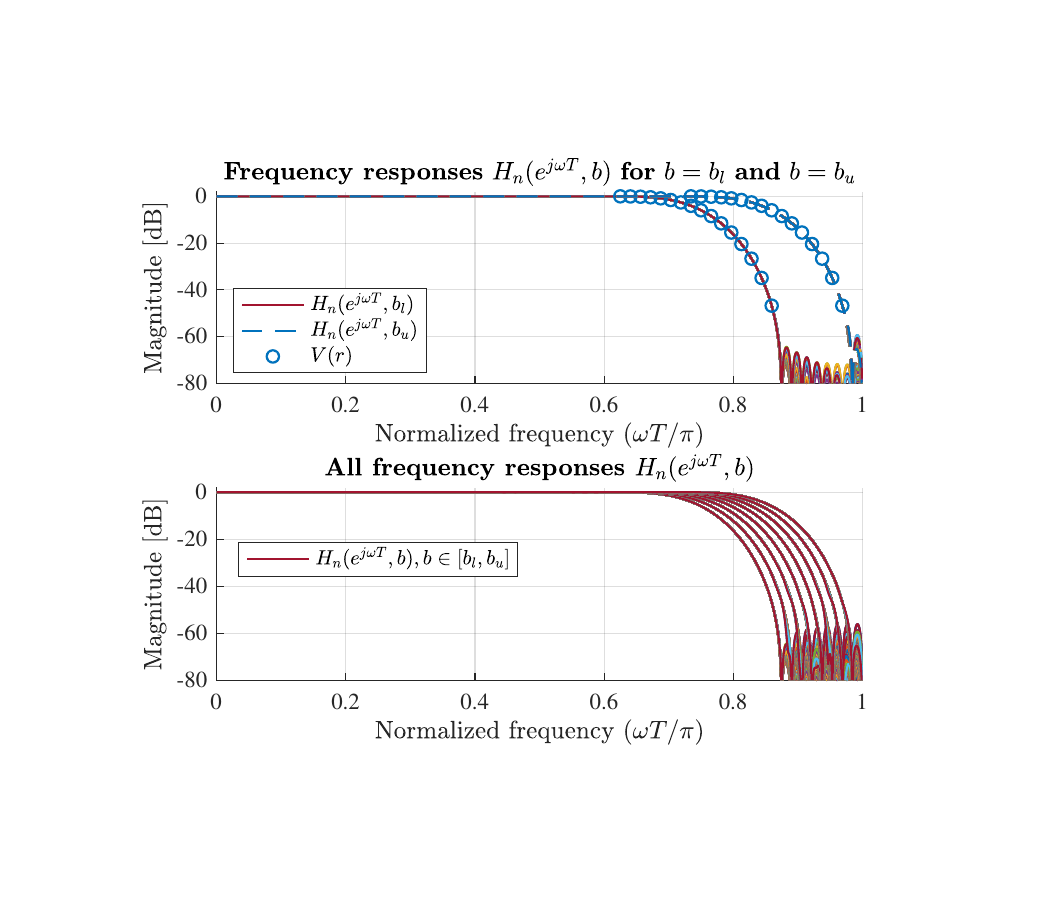}
	\caption{Modulus of the VBW filter transition band samples $V(r)$ and frequency responses $H_n(e^{j\omega T}, b)$, $n=0,1,\dots, M-1$, for $b\in[b_l, b_u]$ in Example 1.}
	\label{fig:ex1}
\end{figure}

The VBW filter specification, which exactly matches the frequency bins for a given DFT length, i.e., there is no need to perform rounding as stated in Section~\ref{subsec:procedure_complexity}, is taken from \cite{Moryakova_2024_HFSO} as follows: $\Delta=0.25\pi$, $b\in[96\pi/128, 110\pi/128]$. For a fair comparison of filter design results and overall complexity, the filter length in the proposed method is set equal to that obtained with the method in \cite{Moryakova_2024_HFSO}, which in this case\footnote{In \cite{Moryakova_2024_HFSO}, the value $L=33$ was reported under different convergence tolerance parameters. Here, with smaller tolerance, applying the same algorithm proposed in \cite{Moryakova_2024_HFSO}, we obtained $L=31$.} is $L=31$.
The DFT length is $N=128$ and $M=98$.
Thus, the transition band and bandwidth in terms of bins are $\Delta_N=16$ and $b_N(b)\in[b_{N}(b_l), b_{N}(b_u)]=[48,55]$, respectively. 

The magnitude responses of the VBW filter designed using the proposed method are presented in Fig. \ref{fig:ex1}. The upper plot shows the transition-band coefficients $V(r)$, which are the same for different values of $b$, together with the time-invariant magnitude responses $H_n(e^{j\omega T}, b)$ for $n=0,1,\dots, M-1$ and $b=[b_l, b_u]$, whereas the lower plot illustrates the magnitude responses for all values of the variable parameter within the given range, i.e., $b\in[b_l, b_u]$. Note that $b_N(b_l)=b_l$ and $b_N(b_u)=b_u$ in this example. The implementation and reconfiguration complexities per output sample along with the stopband maximum level (SBML) and stopband energy (SBE) averaged among all the responses $H_n(e^{j\omega T}, b)$ are listed in Table \ref{tab:example1}. It is clearly seen that the proposed approach allows to significantly reduce the implementation and online reconfiguration complexity compared to the filters designed in the TD, but requires more memory compared to the TD implementation in \cite{Lowenborg_2006}.
In comparison with \cite{Moryakova_2024_HFSO}, the implementation and reconfiguration complexities are the same, but the offline design complexity is reduced. Additionally, although the proposed LS approach shows degradation in the SBML compared to all the methods, which are based on minimax optimization, the SBE averaged among all the responses $H_n(e^{j\omega T}, b)$ is the lowest one, as expected for a LS design.

\begin{table*}[t]
	\caption{Overall Complexity per Output Sample and Approximation Error for the VBW Filter in Example 1}
	\begin{center}
		\begin{tabular}{|l| c c c c |c c c| c c c| c c|}
			\hline 
			\multirow{2}{*}{\textbf{Designed/Implemented}}&\multicolumn{4}{c|}{\textbf{Implementation parameters}} &\multicolumn{3}{c|}{\textbf{Implementation complexity}}&\multicolumn{3}{c|}{\textbf{Reconfiguration complexity}}&\multicolumn{2}{c|}{\textbf{Approximation error}}\\
			& $L$ & $N$ & $M$ & $P$ & $R_{mf}$ & $R_{mv}$ & $R_{a}$ &$R_{md}$&$R_{ad}$& Mem.& SBML, dB & SBE, dB \\
			\hline \hline
			TD/TD \cite{Lowenborg_2006} & 29 & - & - & 4 & 75 & 4 & 145 &0&1&1&$-61.9$&$-78.3$\\
			TD/FD \cite{Moryakova_2023} & 29 & 128 & 100 & - & 5.2 & 1.9 & 23.8 &5.2&5.1&640&$-61.9$&$-78.3$\\	
			FD/FD \cite{Moryakova_2024_HFSO} & 31 & 128 & 98 & - & 5.3 & 0.3 & 21 &0&0&15& $-61.2$& $-83.6$ \\
			\hline
			Proposed & 31 & 128 &98 & - & 5.3 & 0.3 & 21 &0&0&15&$-56.1$  &$-89.0$ \\
			\hline
		\end{tabular}
		\label{tab:example1}
	\end{center}
\end{table*}
\begin{table*}[t]
	\caption{Overall Complexity per Output Sample and Approximation Error for the VBW Filter in Example 3}
	\begin{center}
		\begin{tabular}{|l| c c c c |c c c| c c c| c c|}
			\hline 
			\multirow{2}{*}{\textbf{Designed/Implemented}}&\multicolumn{4}{c|}{\textbf{Implementation parameters}} &\multicolumn{3}{c|}{\textbf{Implementation complexity}}&\multicolumn{3}{c|}{\textbf{Reconfiguration complexity}}&\multicolumn{2}{c|}{\textbf{Approximation error}}\\
			& $L$ & $N$ & $M$ & $P$ & $R_{mf}$ & $R_{mv}$ & $R_{a}$ &$R_{md}$&$R_{ad}$& Mem.& SBML, dB & SBE, dB \\
			\hline \hline
			TD/TD \cite{Lowenborg_2006} & 27 & - & - & 4 &  70& 4 &135  &0&1&1&$-63.1$&$-81.3$\\
			TD/FD \cite{Moryakova_2023} & 27 & 128 & 102 & - &  $5.1$& $1.9$ & $23.3$ &$5.0$&$5.0$&$640$&$-63.1$&$-81.3$\\	
			FD/FD \cite{Moryakova_2024_HFSO} & 31 & 128 & 98 & - & 5.3 & 0.3 & 21 &0&0&15& $-61.2$& $-86.9$ \\
			\hline
			Proposed & 31 & 128 &98 & - & 5.3 & 0.3 & 21 &0&0&15&$-56.1$  &$-92.2$ \\
			\hline
		\end{tabular}
		\label{tab:example3}
	\end{center}
\end{table*}

\subsection{Example 2}
\begin{figure}[t]
	\includegraphics[trim={2.4cm 2.6cm 2.4cm 2.6cm}, clip,width=\linewidth]{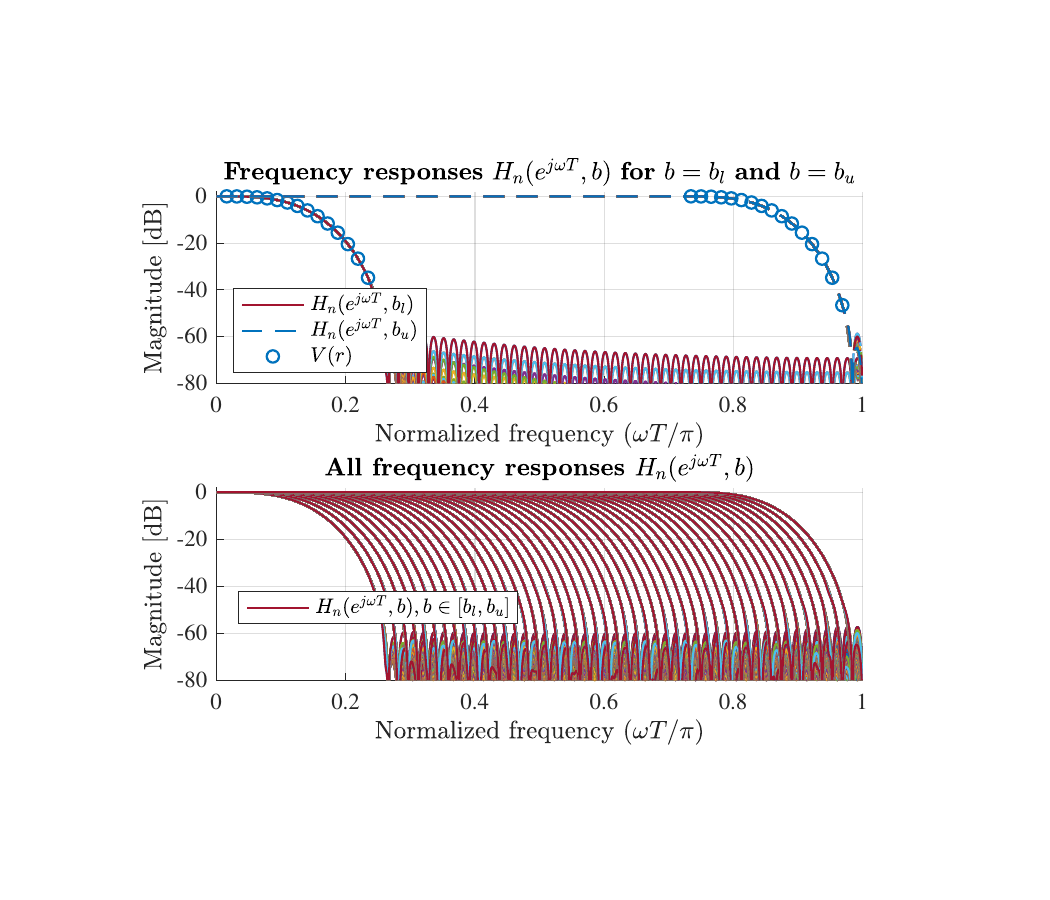}
	\caption{Modulus of the VBW filter transition band samples $V(r)$ and frequency responses $H_n(e^{j\omega T}, b)$, $n=0,1,\dots, M-1$, for the entire variable band in Example 2.}
	\label{fig:ex2}
\end{figure}

This example studies a VBW filter allowing to vary the bandwidth in a wide range without an increase in complexity. The specification of the VBW filter is taken from Example~1 but the variable parameter varies within a wider range, specifically, $b\in[16\pi/128, 110\pi/128]$. 

The time-invariant magnitude responses of the VBW filter designed using the proposed method are shown in Fig. \ref{fig:ex2}. After the design, the SBML among all $M$ responses and all $b$-values is $-57.4$~dB while the SBE is $-88$~dB. Since the same set of the transition band values $V(r)$ is used for different values of $b$, the implementation and reconfiguration complexities of the obtained filter in this case is the same as in Example~$1$, while the complexities for \cite{Lowenborg_2006} and \cite{Moryakova_2023} will be much higher for this wide range of the variable parameter $b$. This makes the proposed approach highly attractive when the entire range of the VBW must be covered, as is also the case in \cite{Moryakova_2024_HFSO}.

\subsection{Example 3}
In this example, we consider a VBW filter specification that does not match the DFT frequency bins, i.e., rounding of the specification needs to be performed, as described in Section~\ref{subsec:procedure_complexity}. 

Based on the specification given by 
$\Delta=0.27\pi$, $b\in[0.76\pi, 0.85\pi]$, the estimated filter order is ${N}_D=26$ and the corresponding DFT length is $N=128$. Further, the transition width $\Delta_N$ is computed and reduced from $17$ to $16$, that corresponds to $\Delta_D=0.25\pi$. 
Thus, after rounding, the new specification for the design is tighter compared to the initial one, specifically it is the same as in Example~$1$, that gives $L=31$ (for the same comparison reason as in Example~$1$), $N=128$, $M=98$, $b_N(b_l)=48$, and $b_N(b_u)=55$ (that corresponds to $\tilde{b}_l=96\pi/128$ and $\tilde{b}_u=110\pi/128$).	

Table \ref{tab:example3} lists the overall complexity per output sample together with the approximation errors of the proposed method in comparison with existing approaches. It is worth noting that tightening (rounding) the specification has been performed only for the VBW filters designed based on frequency sampling, i.e., for the proposed approach and for \cite{Moryakova_2024_HFSO}, while the filters designed using the methods in  \cite{Lowenborg_2006} and \cite{Moryakova_2023} are based on the initial specification, that gives a slightly smaller filter order and length, correspondingly.
Even though this results in relaxed requirements for \cite{Lowenborg_2006} and \cite{Moryakova_2023}, the complexity savings of the proposed method are still significant and only slightly smaller than when the specification matches the DFT bins.

\subsection{Example 4}
\begin{figure}[t]
	\centering
	\subfloat[LS cost function in \eqref{eq:loss_fun}.]{
		\includegraphics[trim={0.3cm 5cm 0.3cm 5cm},clip,width=\linewidth]{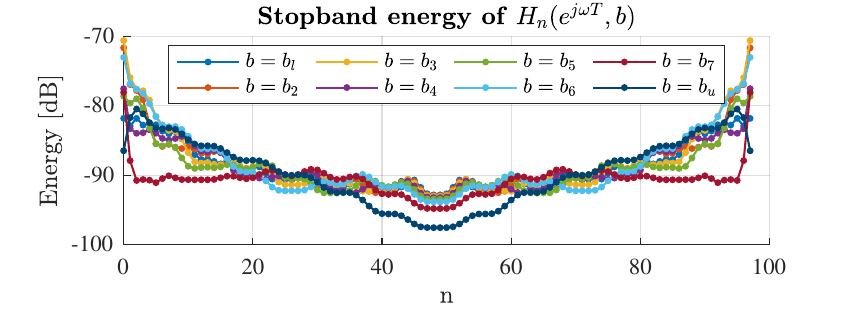}%
		\label{fig:ex4b}
	} \\[1ex]
	\subfloat[Weighted LS cost function in \eqref{eq:loss_fun_weighted}.]{
		\includegraphics[trim={0.3cm 5cm 0.3cm 5cm},clip,width=\linewidth]{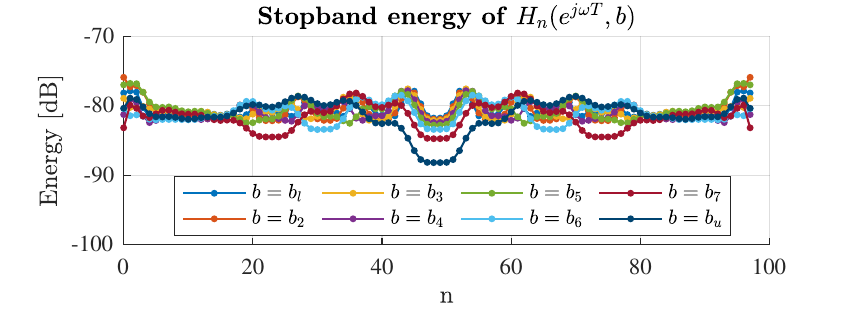}%
		\label{fig:ex4c}
	}
	\caption{Stopband energy of the responses $H_n(e^{j\omega T}, b)$  designed using LS methods in \eqref{eq:loss_fun}  and \eqref{eq:loss_fun_weighted} for $n=0,1,\dots, M-1$ and $b$ varying from $b_l$ to $b_u$.}
	\label{fig:ex4}
\end{figure}

This example studies properties of the set of $M$ time-invariant responses of the $M$-periodic time-varying system. The same specification as in Example~$1$ is considered here.

Due to the circular properties of the impulse responses $d_n(q,b)$, the responses $h_n(q,b)$ have a special structure as discussed in Section~\ref{subsec:VBW_filt}. 
This results in symmetric properties of the their frequency responses, specifically, the stopband energy of $H_n(e^{j\omega T}, b)$ is symmetric w.r.t. $n=(M-1)/2$. This is illustrated in Fig.~\ref{fig:ex4b} for $n=0,1,\dots,M-1$ and $b\in[b_l, b_u]$, where it is also clearly seen that the energy for small $n$ is greater compared to the $n$-values around $(M-1)/2$. This effect can be mitigated by introducing a weighting factor in the LS design method, i.e., the function in  \eqref{eq:loss_fun} can be modified as
\begin{multline}
	E = \sum_{b=b_N(b_l)}^{b_N(b_u)} \sum_{n=0}^{M-1} \frac{1}{2\pi}\int_{\Omega(b)}\big|W_n(b)\\
		\times\big(H_n(\exppjwT{},b)-H_{D_2}(\exppjwT{},b)\big)\big|^2 d(\omega T)
	\label{eq:loss_fun_weighted}
\end{multline}
with weights $W_n(b)$ for $n=0,1,\dots,M-1$ and  $b\in[b_l, b_u]$, which can be obtained based on the stopband energy of the corresponding responses $H_n(e^{j\omega T},b)$ using the design in \eqref{eq:loss_fun}. 
The weighted approach leads to a more flat distribution of the energy values among different responses $H_n(e^{j\omega T}, b)$ as illustrated in Fig.~\ref{fig:ex4c}. 
Here, the maximum SBE level reduced from $-70.8$ to $-75.9$ dB while the averaged among all responses energy increased from $-89.0$ dB to $-80.1$ dB. 

\section{Conclusion}
\label{sec:conclusion}
This paper introduced a closed-form LS design approach for FC-based VBW filters. The proposed design method allows not only closed-form design with substantial complexity reductions and minimized stopband energy of an LPTV system frequency responses compared to minimax-optimization based approaches for a given performance, but also allows to eliminate the dependence of the implementation and online reconfiguration complexities on the variable bandwidth range. Additionally, it was demonstrated that the maximum error energy over the whole set of the time-invariant filters of the LPTV system can be reduced by incorporating weighting functions in the LS design.


\bibliographystyle{IEEEtran}
\bibliography{bibliography}

\vfill\pagebreak

\end{document}